\documentclass[twocolumn,showpacs,aps,citeautoscript,superscriptaddress]{revtex4-2}
\usepackage[utf8]{inputenc}
\usepackage[T1]{fontenc}
\usepackage{mathtools}
\usepackage{tikz}
\usepackage{csvsimple}
\usepackage{booktabs}
\usepackage{amsfonts}
\usepackage{gensymb}
\usepackage{amssymb}
\usepackage{graphicx}
\usepackage{siunitx}
\usepackage{amsmath}
\usepackage{comment}
\usepackage{bm}
\usepackage{color}
\usepackage{textgreek}
\usepackage{subfig}
\usepackage{multirow}
\usepackage{booktabs}
\usepackage{numprint}
\usepackage{rotating}
\usepackage{mhchem}
\usepackage{ragged2e}
\usepackage{breqn}
\usepackage{braket}
\usepackage{hyperref}
\usepackage{dcolumn}
\usepackage{float}
\usepackage{placeins}
\usepackage{tabularx}
\usepackage[para,online,flushleft]{threeparttable}
\newcommand{\sout}[1]{\sout{#1}}

\begin{document}

%
\title{Genuine Molecular Effects in Bound-State Quantum Electrodynamics}

\author{Kjell Janke}
\affiliation{\Philipps}

\author{Konstantin Gaul}
\email[Electronic mail: ]{konstantin.gaul@uni-mainz.de}
\affiliation{\HIM}
\affiliation{\JGU}
\affiliation{\GSI}

\author{I.~Agust\'{\i}n Aucar}
\email[Electronic mail: ]{agustin.aucar@conicet.gov.ar}
\affiliation{\IMIT}
\affiliation{\RUG}

\author{Karol Kozio{\l}}
\affiliation{\NCBJ}

\author{Gustavo A.~Aucar}
\affiliation{\IMIT}

\author{Robert Berger}
\email[Electronic mail: ]{robert.berger@uni-marburg.de}
\affiliation{\Philipps}

\newcommand{\Philipps}{
Fachbereich Chemie, Philipps–Universit\"at Marburg, 
35032 Marburg, Germany
\vspace*{0.1cm}
}

\newcommand{\HIM}{
Helmholtz Institut Mainz, 55099 Mainz, Germany
\vspace*{0.1cm}
}
\newcommand{\GSI}{GSI Helmholtzzentrum für Schwerionenforschung GmbH, 64291 Darmstadt, Germany
\vspace*{0.1cm}}
\newcommand{\JGU}{Institut f\"ur Physik, Johannes Gutenberg-Universit\"at Mainz, 55099 Mainz, Germany
\vspace*{0.1cm}}

\newcommand{\NCBJ}{
Narodowe Centrum Bada\'{n} J\k{a}drowych (NCBJ), 
05-400 Otwock-\'{S}wierk, Poland
\vspace*{0.1cm}
}

\newcommand{\IMIT}{
Instituto de Modelado e Innovaci\'on Tecnol\'ogica (UNNE-CONICET),
Universidad Nacional del Nordeste, 
W3404AAS Corrientes, Argentina
\vspace*{0.1cm}
}

\newcommand{\RUG}{
Van Swinderen Institute for Particle Physics and Gravity, University of Groningen, 9747 AG Groningen, The Netherlands
\vspace*{0.1cm}
}

\date{\today}


\begin{abstract}
Quantum electrodynamics (QED) currently features prominently at the precision frontier in describing the structure of atoms and molecules. Dominant QED effects are governed by the electronic structure in the vicinity of the nucleus and are therefore generally regarded as atomic in origin, with molecular contributions largely arising from additive atomic contributions. By studying the nuclear magnetic resonance shielding tensors of thallium in Tl$X$, we reveal genuinely molecular QED effects that strongly depend on the bonding partner ($X$ = F, Cl, Br, I, At) and are absent in the Tl$^+$ ion.
\end{abstract}

\maketitle

\textit{Introduction}.---Quantum electrodynamics (QED) is the prototype of an extremely well-tested quantum field theory and an integral part of the Standard Model of particle physics. QED is nowadays understood to be ultimately required for the most accurate descriptions of electronic structures in atoms and molecules, when only electromagnetic interactions are considered \cite{indelicato:1990,indelicato:1992,sunnegren:1998,shabaev:1998,shabaev:2001,sapirstein:2001,romero:2002pra,romero:2002ijms,sapirstein:2003,liu:2003,jentschura:2006,indelicato:2007,aucar:2014,pasteka:2017,ginges:2017,koziol:2018,wehrli:2021,wehrli:2022}. Leading-order QED contributions in stable atoms and molecules arise from vacuum polarization (VP) and electron self-energy (SE), which are most pronounced near the atomic nuclei and for inner-shell electrons. Consequently, QED effects are perceived to be mostly atomic in nature, and models that just add up individual atomic contributions are likely to describe the major effects.

We recently embarked on the study of QED effects on molecular properties, and later proposed an atomic-based model to account for the atomic-like contributions of QED to nuclear magnetic resonance (NMR) parameters \cite{koziol:2019,colombo:2022}. We subsequently applied this model to investigate QED effects on the NMR shielding tensors of the diatomic thallium halides Tl$X$, with $X$ = F, Cl, Br, I, At \cite{koziol:2024,koziol:2024_erratum}. These molecules have been considered for precision studies of fundamental physics, including searches for symmetry-violating interactions~\cite{hinds:1980,schropp:1987,cho:1991,hinds:1997,demille:2024,centrex:2025}. Within this atomic-based model, only a very weak molecular trend is observed across the Tl$X$ series, with the QED effects being nearly independent of the bonding partner $X$. With the recent molecular approach for treating QED effects in a two-component (2C) quasi-relativistic framework implemented by some of the authors of this work, we have demonstrated excellent agreement with reference data from four-component (4C) relativistic studies of QED effects on the energy levels of atoms and diatomic molecules, when the same prescriptions for the leading-order QED corrections are used in both frameworks \cite{janke:2025}. Using a linear-response approach, we also showed that orbital energy corrections obtained from perturbative and self-consistent treatments of QED contributions are in close agreement. These highly encouraging results motivated us to consider, in the 2C and 4C frameworks, also QED corrections to magnetic response properties in atoms and molecules, particularly NMR shielding tensors.

At variance with recent models for QED in molecules which exploit the atomic nature of these corrections \cite{koziol:2019,colombo:2022,koziol:2024,koziol:2024_erratum}, we observe, by systematically varying the bonding partner, the unexpected emergence of genuinely molecular QED effects (clearly absent in atomic systems) in the 4C and 2C frameworks. These effects are beyond known additive atomic effects, such as bond length elongation  \cite{sunaga:2022}. We briefly describe our theoretical methodology in the next paragraph, followed by the main computational settings, and proceed to the main results of our study and their discussion.

\textit{Theory}.---The NMR shielding tensor $\bm{\sigma}(K)$, of nucleus $K$ with magnetic moment $\vec{\mu}_{K}$ in a uniform external magnetic field $\vec{B}$, is obtained from the variationally stationary electronic energy $E(\vec{B},\vec{\mu}_K)$ of an atom or molecule, and its elements are
\begin{equation}
\sigma_{ij}(K)=\left.\frac{\partial^2 E(\vec{B},\vec{\mu}_K)}{\partial\mu_{K,i}\,\partial B_{j}}\right|_{\vec{B}=\vec{\mu}_K=0} ,
\label{eqgeneralshieldingtensor}
\end{equation} 
where $i$ and $j$ are its Cartesian components. As for other electric or magnetic second-order atomic or molecular properties, this tensor can be evaluated using linear-response functions within either non-relativistic (NR) or relativistic polarization propagator theory \cite{oddershede:1975,aucar:1993,aucar:1999}. In the NR and 2C relativistic formulations, additional expectation-value contributions must also be considered.
The 2C zeroth-order regular approximation (ZORA) and the 4C implementations of this tensor were discussed in Refs.~\cite{saue:2003,gaul:2020,colombojofre:2022,koziol:2024}.
The analysis of the individual electronic mechanisms that yield the largest molecular QED contributions reported in the present work will be described in more detail in Ref.~\cite{janke:tobepublished}.
Herein, we compare the total effects using 2C ZORA and 4C Dirac--Coulomb (DC) calculations.

Furthermore, we employ the relativistic generalization of the NR Ramsey--Flygare relationship, introduced by some of the authors of the present work in Ref.~\cite{aucar:2016b}, between the shielding tensor $\bm{\sigma}(K)$ and the nuclear spin-rotation (NSR) tensor $\mathbf{M}(K)$, which is a genuinely molecular property, to investigate the electronic origin of the atomic-type and molecular-type QED effects separately. Using the so-called M-V model, the shift in the absolute NMR shielding of a nucleus in a molecule relative to its value in the isolated atom can be expressed in terms of the NSR tensors
as \cite{aucar:2016b,aucar:2019}
\begin{align}
 \Delta\bm{\sigma}(K) & = \bm{\sigma}(K)- \bm{1}_{3\times3} \, \sigma^{\mathrm{atom}}(K) \nonumber \\ 
 & \approx \bm{\sigma}^{\mathrm{NSR}}(K) + \Delta\bm{\nu}^{\mathrm{S}}(K) .
\label{eqm5decomposition}
\end{align}
This relation holds only when the gauge origin of the magnetic vector potential is set at the system's center of mass. Here, $\sigma^{\mathrm{atom}}(K)$ is the shielding constant of nucleus $K$ in the corresponding atomic reference system, which for Tl is Tl$^+$, and $\bm{1}_{3\times3}$ denotes the rank-2 identity tensor. Furthermore, for linear molecules, $\bm{\sigma}$ and $\bm{\mathrm{M}}$ are diagonal tensors, with $\bm{\mathrm{M}}$ having two equal nonzero components perpendicular to the molecular axis and a vanishing component parallel to it. Consequently, the nonzero tensor elements of $\bm{\sigma}^{\mathrm{NSR}}(K)$ are
\begin{equation}
\sigma_{\perp}^{\mathrm{NSR}} (K) = \frac{m_{\mathrm{p}} \, J}{\hbar^2 m_{\mathrm{e}}\,g_K} \, {\mathrm{M}}_{\perp}(K) , 
\end{equation}
with $m_{\mathrm{p}}$ and $m_{\mathrm{e}}$ denoting the proton and electron rest masses, respectively, $\hbar$ the reduced Planck constant, $J$ the moment of inertia about the center of mass, $g_K$ the nuclear g-factor, and ${\mathrm{M}}_\perp(K)$ the NSR constant of nucleus $K$, expressed in energy units \cite{aucar:2012}. The remaining term, $\Delta\bm{\nu}^{\mathrm{S}}(K) = \bm{\nu}^{\mathrm{S}}(K) - \bm{1}_{3\times3}\,\nu^{\mathrm{S,atom}}(K)$, is defined in terms of the quantity $\bm{\nu}^{\mathrm{S}}$, which is evaluated as
\begin{equation}
\bm{\nu}^{\mathrm{S}}(K) = \frac{\mu_0}{4\pi} \frac{e^2}{2m_{\mathrm{e}}} \,\mathrm{Re}\left(\biggl\langle\!\!\biggl\langle \frac{\vec{r}_K \times c \vec{\bm{\alpha}}}{|\vec{r}_K|^3} ; \vec{\bm{S}}_{\mathrm{e}} \biggr\rangle\!\!\biggr\rangle_{\omega=0} \right) .
\end{equation}
This zero-frequency ($\omega=0$) linear-response function was introduced in Ref.~\cite{aucar:2016b} as the relativistic generalization of a spin-orbit coupling contribution appearing in an earlier semi-relativistic model \cite{aucar:2016a,aucar:2019}. Here, $\mu_0$ is the permeability of free space, $e$ is the elementary charge, $c$ is the speed of light in vacuum, $\vec{r}_{K} = \vec{r} - \vec{R}_{K}$ is the position vector relative to nucleus $K$, $\vec{R}_{K}$ is the position vector of nucleus $K$, and $\vec{\bm{\alpha}}$ denotes the vector of $4 \times 4$ Dirac matrices. The operator $\vec{\bm{S}}_{\mathrm{e}}=(\hbar/2) \, \bm{1}_{2\times2} \otimes \vec{\bm{\sigma}}^{\mathrm{P}}$ is defined in terms of the Kronecker product between the $2\times 2$ identity matrix and the Pauli matrices, the latter denoted by $\vec{\bm{\sigma}}^{\mathrm{P}}$ to distinguish them clearly from the symbol used for NMR shielding tensors in this work. Compared with Ref.~\cite{aucar:2016b}, the quantity $\bm{\nu}^{\mathrm{S}}$ defined here in SI units corresponds to $1/(2c) \, \bm{\nu}^{\mathrm{S}}$ in the notation of Ref.~\cite{aucar:2016b}, where Gaussian-cgs electromagnetic conventions are employed. The inclusion of $\bm{\nu}^{\mathrm{S}}$ in the calculation of NMR shielding tensors via Eq.~\eqref{eqm5decomposition} yields highly accurate results for light- and heavy-element-containing diatomic and polyatomic molecules \cite{aucar:2016b,aucar:2019,bajac:2021}.

\begin{figure}[!t]
\centering
\includegraphics[width=0.9\linewidth]{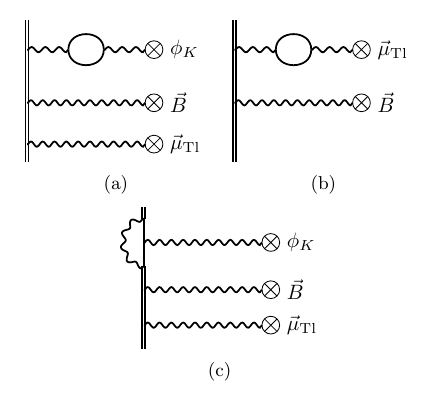}
\caption{\justifying
Feynman diagrams for
(a) the Uehling VP correction to the nuclear electric scalar potential $\phi_K$,
(b) the Schneider~\textit{et al.}~VP correction to the nuclear magnetic vector potential, and
(c) the Flambaum--Ginges effective SE correction associated with the nuclear electric scalar potential.
These diagrams illustrate the leading-order QED corrections considered here for the NMR shielding of a bound electron in the presence of the nuclear magnetic moment $\vec{\mu}_{\text{Tl}}$ and the external magnetic field $\vec{B}$. Double lines represent bound electrons, single lines represent free electrons, and wavy lines denote electromagnetic interactions, including virtual-photon propagators and external-field couplings.
}
\label{figfeynmandiagrams}
\end{figure}

We include leading-order QED corrections through effective potentials. The corresponding Feynman diagrams for the NMR shielding tensors are shown in Fig.~\ref{figfeynmandiagrams}. For the NSR constants and $\Delta\bm{\nu}^{\mathrm{S}}$ terms, the three diagrams are obtained by replacing $\vec{B}$ with the corresponding operator. In all cases, we account for the three possible permutations of $\phi_K$, $\vec \mu_{\text{Tl}}$, and $\vec B$, or the corresponding operator. The interactions represented by the diagrams in Figs.~\ref{figfeynmandiagrams}(a) and \ref{figfeynmandiagrams}(c) were included in the 4C DC relativistic calculations of all three properties (with $\vec{B}$ replaced by the corresponding operator) and in the 2C ZORA calculations of the shieldings, whereas the interaction represented by the diagram in Fig.~\ref{figfeynmandiagrams}(b) was included only in the 2C ZORA calculations of the shieldings.

The leading-order electric VP correction to all three properties (i.e., Fig.~\ref{figfeynmandiagrams}(a) for the case of NMR shieldings) is represented by the Uehling potential \cite{uehling:1935}. 
Higher-order corrections, such as the Wichmann--Kroll term \cite{wichmann:1956}, are not included in the present work. 
In addition, this work includes, only at the 2C ZORA level, besides the electric Uehling potential for VP, the magnetic-loop (ml) contribution to the magnetic vector potential generated by a point-like nuclear magnetic dipole moment (Fig.~\ref{figfeynmandiagrams}(b)), described by Schneider~\textit{et al.}~\cite{schneider:1994},
\begin{align}
\vec{A}^{\mathrm{\,VP,ml}}_K(\vec{r}\,) = &\frac{2\alpha}{3\pi} \vec{A}_{K}(\vec{r}\,) \int_{1}^{\infty}
\left(1-\frac{1}{t^2} \right)^{1/2} \left( 1+\frac{1}{2 t^2} \right) \frac{1}{t} \nonumber \\
& \times\left( \frac{2 |\vec{r}_K| t}{\lambdabar} + 1 \right)
    \exp{\left( -\frac{2 |\vec{r}_K| t}{\lambdabar} \right)} \mathrm{d}t,
\label{eqgreinerpotential}
\end{align}
where $\alpha$ is the fine-structure constant, $\lambdabar=\hbar/(m_{\mathrm{e}}c)=\alpha \,a_0$ the reduced Compton wavelength of the electron, $a_0$ the Bohr radius, $\vec{A}_{K}(\vec{r})=\frac{\mu_0}{4\pi}\frac{\vec{\mu}_{K}\times \vec{r}_K}{|\vec{r}_K|^3}$ the magnetic vector potential generated at position $\vec{r}$ by the point-like magnetic moment $\vec{\mu}_K = e \, g_K / (2 m_{\text{p}}) \, \vec{I}_K$ of nucleus $K$ located at $\vec{R}_{K}$, and $\vec{I}_K$ the spin of nucleus $K$. We note that we intentionally omit a factor of 1/2 in the first parenthesis of Eq.~\eqref{eqgreinerpotential} compared to Ref.~\cite{schneider:1994}, because it appears to be a typographical error, also noted in Refs.~\cite{labzowsky:1997, beier:2000, artemyev:2001, yerokhin:2012, hasted:2024}.
Our current implementation of the magnetic-loop contribution at the 2C ZORA level closely follows that of the SE potentials reported in Ref.~\cite{janke:2025}, and will be described in more detail in Ref.~\cite{janke:tobepublished}.

For the leading-order SE correction (Fig.~\ref{figfeynmandiagrams}(c)), the effective potentials by Flambaum and Ginges \cite{flambaum:2005} are used at both the 4C DC and 2C ZORA levels. The SE vertex corrections (see Ref.~\cite{yerokhin:2012}), however, are not included in the present work. The implementation of the Uehling and Flambaum--Ginges potentials, their evaluation within the 4C and 2C ZORA methods, and the comparison of 2C ZORA and 4C DC calculations at the Hartree--Fock and Kohn--Sham levels are described in Refs.~\cite{sunaga:2022,janke:2025}.

The QED corrections to the shielding tensors of the Tl nucleus in Tl$X$ and in the Tl$^{\mathrm{+}}$ reference system are obtained at both the 4C DC and 2C ZORA levels as
\begin{equation}
    \label{eqsigmadiff}
    \bm{\sigma}^{\mathrm{QED}}(\mathrm{Tl}) = \tilde{\bm{\sigma}}(\mathrm{Tl}) - \bm{\sigma}(\mathrm{Tl}) ,
\end{equation}
whereas those to the NSR and $\bm{\nu}^{\mathrm{S}}$ terms are obtained only at the 4C DC level, as
\begin{align}
\bm{\sigma}^{\mathrm{NSR,QED}}(\mathrm{Tl}) &= \tilde{\bm{\sigma}}^{\mathrm{NSR}}(\mathrm{Tl}) - \bm{\sigma}^{\mathrm{NSR}}(\mathrm{Tl}) ,
\label{eqM-diff} \\
\bm{\nu}^{\mathrm{S,QED}}(\mathrm{Tl}) &= \tilde{\bm{\nu}}^{\mathrm{S}}(\mathrm{Tl}) - \bm{\nu}^{\mathrm{S}}(\mathrm{Tl}).
\label{eqnuS-diff}
\end{align}
Here, $\tilde{\bm{{\sigma}}}(\mathrm{Tl})$, $\tilde{\bm{\sigma}}^{\mathrm{NSR}}(\mathrm{Tl})$, and $\tilde{\bm{\nu}}^{\mathrm{S}}(\mathrm{Tl})$ denote the tensors obtained from calculations in which the electric scalar leading-order QED corrections, namely the Uehling and Flambaum--Ginges potentials, are included variationally. The differences between the corresponding perturbed and ordinary tensors in Eqs.~\eqref{eqsigmadiff}, \eqref{eqM-diff}, and \eqref{eqnuS-diff} are considered to be approximately equal to the third-order perturbative (or quadratic-response) contributions.

In the 2C formulation, the magnetic vector potential of Eq.~\eqref{eqgreinerpotential} is additionally incorporated into the hyperfine operator. This yields the contributions $\bm{\sigma}^{\mathrm{ml}}(\mathrm{Tl})$, which are evaluated using the resulting orbitals.

2C ZORA calculations were performed using a modified \cite{berger:2005, nahrwold:2009, isaev:2012, gaul:2020, bruck:2023, colombojofre:2022, zulch:2022, janke:2025} version \cite{vanwullen:2010} of the program package Turbomole \cite{ahlrichs:1989}. Complex Generalised Kohn--Sham (cGKS) 2C ZORA \cite{chang:1986, lenthe:1993, lenthe:1994, lenthe:1996} orbitals with PBE functional \cite{perdew:1992,perdew:1996}, employing a model potential to alleviate ZORA's gauge dependence as suggested by van~W\"{u}llen \cite{vanwullen:1998}, were converged below a change in the self-consistent field energy of $10^{-10}\,E_{\mathrm{h}}$.
Corresponding 4C DC calculations were performed using DIRAC 25.0~\cite{DIRAC25,dirac-paper}, also employing the PBE functional within the Kohn--Sham approximation, the common gauge origin (CGO) approach, unrestricted kinetic balance (UKB) for generating the small components, and including all (SS|SS) small-component integrals.
The NSR constant and the $\Delta\bm{\nu}^{\mathrm{S}}$ terms, together with their QED corrections, were evaluated only within the 4C DC framework using the same computational settings.
In all 4C calculations, QED effects were incorporated through effective QED potentials for all molecular centres.
Nuclear charge density distributions entering the electron-nucleus electrostatic interaction energy were modeled using spherically symmetric Gaussian functions, such that 
$\rho_{K} \left( \vec{r} \right) = Z_K e \, (\zeta_{K}/\pi)^{3/2} \, \exp{\left(-\zeta_{K} \left| \vec{r}_{K} \right| ^2\right)}$, with $\zeta_{K} = 3/(2 r^2 _{\mathrm{nuc},K})$. $Z_K$ denotes the atomic number of nucleus $K$, and the parameter $r_{\textnormal{nuc},K}$ corresponds to the root-mean-square nuclear radius of nucleus $K$ suggested by Visscher and Dyall \cite{visscher:1997}. Point-like nuclear charge distributions were used for the Uehling potential. The vector potential $\vec{A}_{\text{Tl}}(\vec{r})$ was described using a point-like magnetization distribution in all 4C calculations of NMR shieldings and NSR constants, as well as in the 2C ZORA calculations of NMR shieldings, including those involving Eq.~\eqref{eqgreinerpotential}. Augmented all-electron quadruple-$\zeta$ Dyall basis sets dyall.aae4z \cite{dyall:2006a, dyall:2012a, dyall:2012b, dyall:2023a, dyall:2023b}, were used throughout. Molecular bond lengths of ${}^{205}_{\phantom{0}81}\mathrm{Tl}X$ ($X$ = $^{19}_{\phantom{0}9}$F, $^{35}_{17}$Cl, $^{79}_{35}$Br, $^{127}_{\phantom{0}53}$I, $^{210}_{\phantom{0}85}$At)
in pm are: $r_{\mathrm{TlF}}=208.47$, $r_{\mathrm{TlCl}}=248.48$, $r_{\mathrm{TlBr}}=261.82$, $r_{\mathrm{TlI}}=281.37$, and $r_{\mathrm{TlAt}}=290.71$ \cite{huber:1979,koziol:2024}.
The gauge origin of the external magnetic vector potential was set at the center of mass of the molecules in all calculations, and
the linear-response norm was converged to $10^{-9}$.
\begin{figure}[htbp]
\centering
\subfloat[\centering]{{\includegraphics[width=0.97\columnwidth]{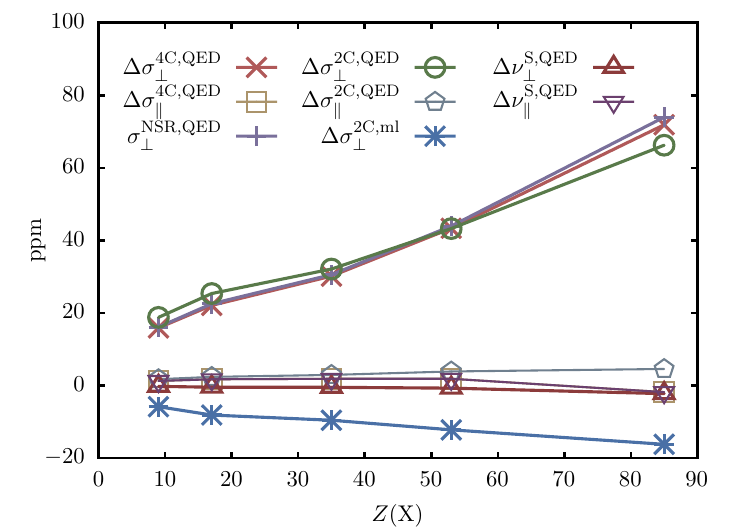}}}\\
\subfloat[\centering]{{\includegraphics[width=0.97\columnwidth]{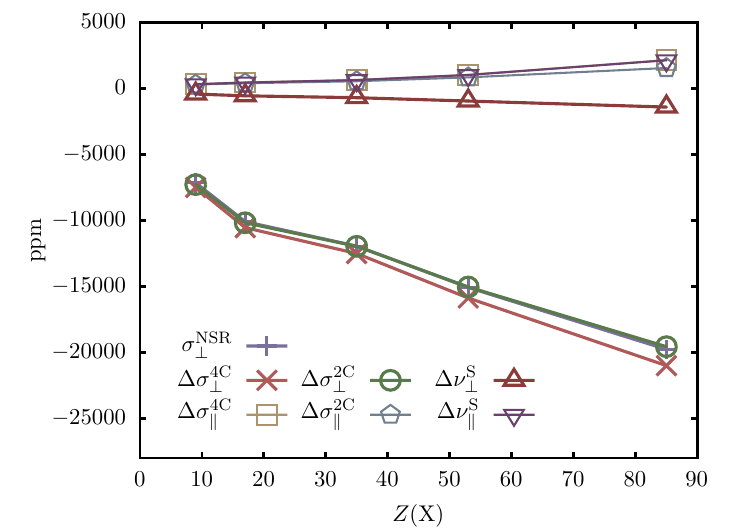}}}
\captionsetup{format=plain,singlelinecheck=false,justification=Justified}
\caption{
\label{comparison_2c_lr_4c_ee_pbe}
Individual tensor elements of the 2C and 4C relative NMR shieldings ($\Delta\bm{\sigma}$) and 4C $\Delta\bm{\nu}^{\mathrm{S}}$, perpendicular ($\perp$) and parallel ($\parallel)$ to the internuclear axis, and 4C NSR constants ($\sigma_\perp^{\mathrm{NSR}}$) for the $^{205}_{\phantom{0}81}$Tl nucleus in Tl$X$ ($X$ = $^{19}_{\phantom{0}9}$F, $^{35}_{17}$Cl, $^{79}_{35}$Br, $^{127}_{\phantom{0}53}$I, $^{210}_{\phantom{0}85}$At), as well as their corresponding QED contributions (without magnetic loop corrections) in ppm, showing strong molecular effects, particularly in the perpendicular components, which acquire a clear nonzero dependence on the nuclear charge number $Z_X$ of the bonding partner $X$. $\Delta\sigma_{\perp}^{\mathrm{2C,ml}}$ denotes the magnetic vector potential contribution of Eq.~\eqref{eqgreinerpotential} and is only evaluated in the 2C ZORA frame.
}
\end{figure}

\textit{Results and discussion}.---For the NMR shielding tensors of Tl in Tl$X$, we find a clear molecular QED trend, i.e., a pronounced dependence of the QED contribution to $\bm{\sigma}({\text{Tl}})$ on the molecular bonding partner $X$ (see Fig.~\ref{comparison_2c_lr_4c_ee_pbe}(a)). The overall trend parallels that observed for the corresponding ordinary shielding values. We find that the molecular QED effect emerges primarily from the QED contribution to the perpendicular component, $\sigma_{\perp}$, whereas the corresponding molecular contribution to the parallel component, $\sigma_{\parallel}$, remains essentially constant along the Tl$X$ series, as shown in Fig.~\ref{comparison_2c_lr_4c_ee_pbe}(a).
In Fig.~\ref{comparison_2c_lr_4c_ee_pbe}, we display the QED contributions to the NMR shieldings of the Tl nucleus in Tl$X$ relative to the Tl$^+$ ion (Fig.~\ref{comparison_2c_lr_4c_ee_pbe}(a)), as well as the corresponding relative NMR shifts, $\Delta \sigma_{\perp} = \sigma_{\perp}-\sigma^{\text{atom}}$, without QED (Fig.~\ref{comparison_2c_lr_4c_ee_pbe}(b)), calculated using both 4C and 2C ZORA methods. While 2C estimates of the absolute NMR shieldings in the perpendicular direction differ from the 4C results \cite{koziol:2024}, the corresponding relative NMR shifts are in good agreement, with deviations of only 3--7\%.
The corresponding QED shifts, $\Delta \sigma_\perp^{\mathrm{QED}}(\mathrm{Tl})$, excluding the magnetic-loop contribution, differ by 19\% and 13\% for the lighter TlF and TlCl, respectively, and by 0--8\% for the heavier TlBr, TlI, and TlAt. To facilitate a direct comparison between the 4C and 2C approaches, the additional magnetic-loop correction of Eq.~\eqref{eqgreinerpotential}, evaluated only within the 2C ZORA framework, is shown separately in Fig.~\ref{comparison_2c_lr_4c_ee_pbe}(a).
It was included in this work not only to consider additional QED contributions but also to exclude the possibility of specific cancellation effects with the Uehling potential.
We find that the magnetic-loop correction is of a similar size to the electric Uehling contribution, and that both have the same sign, a trend also observed in Ref.~\cite{yerokhin:2012} for NMR shieldings in hydrogen-like ions.

For the Flambaum--Ginges SE contributions, we find that, as for energy corrections, they have the opposite sign to the electric VP (Uehling) contributions. In total, however, the SE is larger in magnitude and thus dominates the overall trend of the QED effect. In addition, we find that the molecular QED corrections to the shieldings have the opposite sign to the corresponding total relativistic values without QED effects~\cite{koziol:2024}.
In Fig.~\ref{comparison_2c_lr_4c_ee_pbe}(a), we further demonstrate the genuine molecular origin of the QED effects by showing their contributions to the NSR constants, $\sigma_{\perp}^{\mathrm{NSR}}$, and to the $\Delta\bm{\nu}^{\mathrm{S}}$ terms, both representing inherently molecular quantities.
Applying Eq.~\eqref{eqm5decomposition}, the two terms closely reproduce the 4C ordinary shielding, differing by less than 0.2\%.
In Fig.~\ref{comparison_2c_lr_4c_ee_pbe}(a), we show that the NSR constant captures the main molecular QED contribution, whereas the QED contributions from the $\Delta\bm{\nu}^{\mathrm{S}}$ terms are smaller for the systems investigated here. The nonzero QED contributions to the NSR constants and $\Delta\bm{\nu}^{\mathrm{S}}$ terms, both of which vanish for an isolated atom, provide direct evidence of the genuine molecular origin of the QED effect on the shielding.
Our results are thus supported not only by two molecular program packages, which compare the same QED potentials within 2C and 4C approaches, and by the inclusion of the additional magnetic-loop term of Eq.~\eqref{eqgreinerpotential}, but also by the decomposition of the shielding tensor and the analysis of the NSR tensor. The missing vertex correction to the SE was evaluated for atomic systems in Ref.~\cite{yerokhin:2012} and found to be additive to the correction for heavy atoms ($Z \gtrsim 30$), thereby enhancing the QED effect on NMR shieldings. For light atoms, the correction had the opposite sign and partially canceled the SE contribution. In contrast, no such cancellation was found for the VP, consistent with our results.

\textit{Conclusion}.---We report a genuine molecular QED effect in diatomic thallium halides, Tl$X$ ($X$ = F, Cl, Br, I, At). We investigated the NMR shielding tensors of the Tl nucleus in the Tl$X$ series and support our findings by analyzing the NSR tensors and $\bm{\nu}^{\mathrm{S}}$ terms. Both quantities confirm the molecular origin of the effect, and their sum closely reproduces the relative NMR shift obtained from the two- and four-component calculations. In addition, we included the VP magnetic-loop correction of Eq.~\eqref{eqgreinerpotential} within our two-component ZORA framework, finding an enhancement of the VP contribution and no cancellation.
While additional leading-order QED contributions, such as the SE vertex corrections, are not included in the present work, their inclusion is not expected to lead to a cancellation of the QED effects observed here. Therefore, our present results show a clear molecular QED trend and provide a first estimate of the size of these effects on the shielding tensors of the Tl nucleus in Tl$X$.

\textit{Acknowledgments}---The authors are indebted to Jacinda Ginges, Andr{\'e}~S.~P.~Gomes, Lucas Visscher, and Stefan Buhmann for discussions. We gratefully acknowledge computer time provided by the NHR Center NHR@{}SW at Goethe University Frankfurt and the Center for Information Technology of the University of Groningen.
We thank Goethe University Frankfurt and the University of Groningen for providing access to the Goethe-NHR and Hábrók high-performance computing clusters, respectively.
The NHR computing time is funded by the Federal Ministry of Education and Research, and the state governments participating on the basis of the resolutions of the GWK for national high-performance computing at universities (www.nhr-verein.de/unsere-partner).
K.G. thanks the Fonds der Chemischen Industrie (FCI) for generous funding through a Liebig fellowship.
\bibliographystyle{apsrev4-2}
%

\end{document}